\documentstyle[fortschritte]{article}

\newcommand{\ket}[1]{|\,#1\,\rangle}
\newcommand{\bra}[1]{\langle\,#1\,|}

\newcommand{\NAME}[1]{{#1}, }
\newcommand{\REVIEW}[4]{#1 {\bf #2}, {#3} (#4)}
\newcommand{\BOOK}[4]{{\it #1} (#2, #3, #4)}
\newcommand{\kbar}{\mathchoice{k\hspace{-1.1ex}\raise0.3em\hbox{-}\hspace{0.35ex}}
							  {k\hspace{-1.1ex}\raise0.3em\hbox{-}\hspace{0.35ex}}
							  {k\hspace{-0.9ex}\raise0.17em\hbox{\small -}\hspace{0.25ex}}
							  {k\hspace{-0.8ex}\raise0.15em\hbox{\tiny -}\hspace{0.2ex}}}

\begin{document}

\title{Kicked Rotor in Wigner Phase Space}

\author{M. Bienert, F. Haug, and W.P.~Schleich \\ Abteilung f\"ur Quantenphysik, Universit\"at Ulm, Albert--Einstein--Allee 11, \\ 89069 Ulm, Germany \\ M.G.~Raizen \\ Center for Nonlinear Dynamics and Department of Physics, \\ The University of Texas at Austin, Austin, Texas 78712-1081}

\maketitle

\begin{abstract}
We develop the Wigner phase space representation of a kicked particle for an arbitrary but periodic kicking potential. We use this formalism to illustrate quantum resonances and anti--resonances.
\end{abstract}                                                                                                 
\section{Introduction}
												  For many years the field of quantum chaos \cite{bib:haake,bib:blumel,bib:casati} had been the exclusive play--ground of theoretical physics. However, the recent success in experimentally realizing cold atom sources and the newly emerging field of atom optics \cite{bib:kazan,bib:adams,bib:moore1} are slowly changing this situation in favor of experimental physics. Indeed, a series of atom optics experiments \cite{bib:moore2,bib:raizen,bib:doherty,bib:darcy} has verified fingerprints of quantum chaos such as dynamical localization, quantum resonances, quantum dynamics in a regime of classical anomalous diffusion and accelerator modes. Even the reconstruction of the wave function of the kicked particle in amplitude {\em and} phase seems now to be feasible \cite{bib:bienert}.

Quantum effects are due to quantum interference. Interference phenomena stand out most clearly in the Wigner function description of quantum theory \cite{bib:book}. This feature serves as our motivation to develop in the present paper the Wigner function treatment of a kicked particle. Similar steps have already been taken in \cite{bib:berry,bib:cohen,bib:zurek,bib:habib,bib:gardiner}. However, our motivation and ultimate goal is slightly different: We try to gain a deeper understanding of dynamical localization. In this spirit the purpose of the present paper is to lay the foundations for this long term goal and to summarize the inevitable formalism.

Our paper is organized as follows: In Sec.~\ref{sec:model} we briefly review the essential ingredients of the delta function kicked particle. Here, we do not specify the kicking potential but only assume that it is periodic in space. Moreover, we introduce dimensionless variables which we use throughout the paper. In this way only two dimensionless parameters --- the scaled Planck's constant and the scaled kicking strength --- determine the quantum dynamics of this model. In Sec.~\ref{sec:timeev} we derive the stroboscopic time evolution, that is the maps for the state vector and the Wigner function. We then in Sec.~\ref{sec:resstate} use the state vector map to review the effect of quantum resonances. Section~\ref{sec:reswig} is dedicated to an illustration of this phenomenon and of anti--resonances in Wigner phase space. We conclude in Sec.~\ref{sec:concl} by summarizing our main results. In order to keep the paper self--contained we have included the relevant calculations in three Appendices.

\section{The model}
\label{sec:model}

In the present section we briefly summarize the classical and quantum mechanical model of the kicked particle. In contrast to the standard treatments we do not specify the form of the kicking potential, except that it should be periodic in space. Moreover, we introduce dimensionless variables which reduce the number of parameters. We first consider the motion of a classical particle of mass $M$, characterized by coordinate $\tilde x$ and momentum $\widetilde p$. Then we turn to the quantum description.

The motion of the particle is driven by a sequence of $\delta$--function kicks with period $T$ described by the Hamiltonian
\begin{equation}
{\widetilde H} = \frac{{\widetilde p}^2}{2 M} + \widetilde K \widetilde V{({\tilde x})}\sum_{n=-\infty}^\infty\delta(\widetilde t-nT).
\label{eq:expham}
\end{equation}
The strength of the kick depends on the position of the particle via a potential $\widetilde K\widetilde V(\tilde x)$ where $\widetilde K$ denotes the kick amplitude and $\widetilde V(\tilde x)$ contains the spatial dependence with unit amplitude. Throughout the paper we consider potentials which are periodic with period $\lambda\equiv 2\pi/k_0$. Moreover, we also assume the symmetries $\tilde V(-\tilde x)=-\tilde V(\tilde x)$ and $\tilde V(\tilde x\pm\lambda/2)=-\tilde V(\tilde x)$.

For the further analysis it is convenient to use scaled variables. In particular, we introduce the dimensionless coordinate $x\equiv k_0\tilde x$, momentum $p\equiv (k_0 T/M) \widetilde p$ and time $t\equiv\widetilde t/T$. With these variables the Hamiltonian, Eq.~(\ref{eq:expham}), transforms into
$$
\widetilde H \equiv \frac{M}{k_0^2 T^2} H
$$
where 
\begin{equation}
H \equiv \frac{1}{2}p^2+K V(x)\sum_{n=-\infty}^\infty\delta(t-n)
\label{eq:ham}
\end{equation}
includes the stochasticity parameter $K \equiv  \widetilde K k_0^2 T/M $ and the scaled potential $V(x)\equiv\widetilde V(x/k_0)$ now enjoys the period $2\pi$.

We now turn to the quantum description with position and momentum operators $\hat{\tilde x}$ and $\hat{\tilde p}$. They satisfy the familiar commutation relation 
$[\hat{\tilde x},\hat{\widetilde p}]=i\hbar$ which in dimensionless variables reads
$[\hat x,\hat p]=i\kbar$. Here we have introduced the scaled Planck's constant 
$\kbar\equiv \hbar k_0^2 T/ M$.

Moreover, in these dimensionless variables the Schr\"odinger equation 
$$
i \hbar \frac{\partial}{\partial \widetilde t} \ket\psi = \skew3\hat{\widetilde H}\ket \psi
$$
takes the form 
\begin{equation}
i \kbar \frac{\partial}{\partial t} \ket\psi = \hat H \ket \psi
\label{eq:sgl}
\end{equation}
where we have replaced in the classical Hamiltonian $H$, Eq.~(\ref{eq:ham}), the position and momentum $x$ and $p$ by the corresponding operators giving rise to the quantum mechanical Hamiltonian $\hat H$.

\section{Time evolution}
\label{sec:timeev}

We now turn to the discussion of the time evolution of this kicked system. Here we pursue two different approaches: we first concentrate on the dynamics of the state vector, and  then use these results to derive the time evolution of the Wigner phase space distribution. Due to the stroboscopic behavior of the potential energy we reduce the continuous time evolution to a discrete mapping.

\subsection{State vector}

The Hamiltonian
$$
\hat H \equiv \frac{\hat p^2}{2}+K V(\hat x)\sum_{n=-\infty}^\infty\delta(t-n)
$$
consists of two parts: {\it (i)} The operator of kinetic energy and {\it (ii)} the operator of potential energy which is explicitly time dependent. However, the latter part is only of importance for integer $t$. Between two kicks it vanishes and the state $\ket{\psi_n}$ evolves freely according to 
\begin{equation}
\ket{\psi_n'} = \hat U_{\rm free}(\hat p)\ket{\psi_n}\equiv \exp\left[-i\frac{\hat p^2}{2 \kbar} \right]\ket{\psi_n}.
\label{eq:free}
\end{equation}
Here, we have propagated the state over one time unit $t=1$. 

For integer $t$, the potential energy dominates over the kinetic energy and we can neglect it. This feature allows us to integrate the Schr\"odinger equation, Eq.~(\ref{eq:sgl}), over one kick. 
The state $\ket{\psi_{n+1}}$ immediately after a $\delta$--function kick 
is related to the state $\ket{\psi_n'}$ just before the kick by 
\begin{equation}
\ket{\psi_{n+1}} = \hat U_{\rm kick}(\hat x)\ket{\psi_n'}\equiv \exp\left[-i\frac{K}{\kbar} V(\hat x)\right]\ket{\psi_n'}.
\label{eq:kick}
\end{equation} 
We emphasize that neglecting the kinetic energy is not an approximation since the $\delta$--function only acts at an instant of time with an infinite strength.

When we combine Eqs.~(\ref{eq:free}) and (\ref{eq:kick}) the complete time evolution over one period reads
\begin{equation}
\ket{\psi_{n+1}} = \hat U_{\rm kick}(\hat x)\hat U_{\rm free}(\hat p)\ket{\psi_n} = \exp\left[-i\kappa V(\hat x)\right]\exp\left[-i\frac{\hat p^2}{2 \kbar} \right]\ket{\psi_n}
\label{eq:svmap}
\end{equation}
and maps the state $\ket{\psi_n}$ onto $\ket{\psi_{n+1}}$. Here we have introduced the abbreviation $\kappa\equiv K/\kbar$.

We find the quantum state $\ket{\psi_N}$ after $N$ kicks by applying the Floquet operator
\begin{equation}
\hat U(\hat x, \hat p) \equiv \hat U_{\rm kick}(\hat x)\hat U_{\rm free}(\hat p)
\label{eq:ukuf}
\end{equation}
$N$ times onto the initial state $\ket{\psi_0}$.

Since the potential $V(x)$ is periodic, that is $V(x+2\pi)=V(x)$, the kick operator $\exp\left[-i\kappa V(\hat x)\right]$ is also periodic and we can expand it into Fourier series
\begin{equation}
\hat U_{\rm kick}(\hat x) =e^{-i\kappa V(\hat x)}= \sum_{l=-\infty}^{\infty} S_l\left(\kappa\right) e^{-i l \hat x}
\label{eq:expans}
\end{equation}
with expansion coefficients
\begin{equation}
S_l \left(\kappa\right) \equiv \frac{1}{2 \pi}\int\limits_{-\pi}^{\pi}\! d \xi \, e^{i l \xi}e^{-i\kappa V(\xi)}.
\label{eq:coestate}
\end{equation}
In this Fourier representation the operator nature of the $\hat U_{\rm kick}$ only enters through the Fourier operator $\exp[-il\hat x]$.

With the the help of the relation Eq.~(\ref{eq:expans}) the Floquet operator Eq.~(\ref{eq:ukuf}) takes the form
$$
\hat U(\hat x, \hat p)=\sum_{l=-\infty}^{\infty} S_l\left(\kappa\right) e^{-i l \hat x} \exp\left[-i\frac{\hat p^2}{2 \kbar} \right].
$$

We conclude this subsection by discussing the action of this operator on a momentum eigenstate. For this purpose we calculate the representation of $\hat U$ in the momentum basis by inserting the completeness relation $1\skew{-8} {\rm l}=\int\! dp\, \ket p \bra p$ twice. When we recall the formula 
\begin{equation}
e^{-i l \hat x}\ket p = \ket{p-l\kbar}
\label{eq:shift}
\end{equation} 
we arrive at
\begin{equation}
\hat U = 
\int\limits_{-\infty}^{\infty}\!dp\sum_{l=-\infty}^{\infty} S_l(\kappa) \exp\left[-i\frac{p^2}{2 \kbar}\right]\ket{p-l\kbar}\bra{p}.
\label{eq:floquetopmom}
\end{equation}
Hence, the Floquet operator $\hat U$ couples only momentum eigenstates separated by integer multiples of $\kbar$. 

\subsection{Wigner function}

We now analyze the dynamics of the kicked particle from the point of view of phase space. In particular, we discuss the time evolution of the corresponding Wigner function. Two possibilities offer themselves: {\it (i)} We can solve the quantum Liouville equation \cite{bib:book} of the kicked particle, or {\it (ii)} we can use the mapping of the state vectors derived in the preceding section to obtain a mapping of the corresponding Wigner functions. In the present section we pursue the second approach.

For this purpose we recall the definition
\begin{equation}
W_{n+1}(x,p) \equiv \frac{1}{2\pi\kbar}\int\limits_{-\infty}^\infty \!d\xi\, e^{-i p \xi/\kbar} \langle x+\frac{1}{2}\xi|\psi_{n+1}\rangle\langle\psi_{n+1}|x-\frac{1}{2}\xi\rangle
\label{eq:wigner}
\end{equation} 
of the Wigner function of the state $\ket{\psi_{n+1}}$. 

When we substitute the mapping, Eq.~(\ref{eq:svmap}), into the right-hand side of Eq.~(\ref{eq:wigner}), we arrive at
\begin{equation}
W_{n+1}(x,p) = \frac{1}{2\pi\kbar}\int\limits_{-\infty}^\infty \!d\xi\, e^{-i p \xi/\kbar}
U_{\rm kick}(x+\xi/2) U^\ast_{\rm kick}(x-\xi/2)\langle x+\xi/2|\hat U_{\rm free}|\psi_n\rangle\langle\psi_n|\hat U_{\rm free}^\dagger|x-\xi/2\rangle.
\label{eq:wigstep10}
\end{equation}
Here we have used the property $\hat U_{\rm kick}(\hat x) \ket x = U_{\rm kick}(x) \ket x$.

Since the potential $V(x)$ is periodic, the bilinear form 
$$
U_{\rm kick}(x+\xi/2) U^\ast_{\rm kick}(x-\xi/2)=e^{-i\kappa\left[V(x+\xi/2)-V(x-\xi/2)\right]}\equiv  e^{-i\kappa{\mathcal V}(x,\xi/2)}
$$
with the generalized potential 
\begin{equation}
{\mathcal V}(x,y)\equiv V(x+y)-V(x-y)
\label{eq:defV}
\end{equation}
is also periodic in $\xi/2$ with period $2\pi$. We can therefore expand 
\begin{equation}
e^{-i\kappa{\mathcal V}(x,\xi/2)}=\sum_{l=-\infty}^\infty {\mathcal S}_l(\kappa; x) e^{-i l \xi/2}
\label{eq:wigfouser}
\end{equation}
into a Fourier series where the expansion coefficients
\begin{equation}
{\mathcal S}_l(\kappa; x)\equiv \frac{1}{2 \pi} \int\limits_{-\pi}^{\pi} \!d\left(\frac{\xi}{2}\right) e^{i l \xi/2}  e^{-i\kappa{\mathcal V}(x,\xi/2)}
\label{eq:wigec}
\end{equation}
still depend on the position $x$ and are periodic with a period of $2\pi$.

With the help of the Fourier series, Eq.~(\ref{eq:wigfouser}), the mapping of the Wigner function, Eq.~(\ref{eq:wigstep10}), yields
\begin{equation}
W_{n+1}(x,p) =  \sum_l {\mathcal S}_l(\kappa; x) \, \frac{1}{2\pi\kbar}  \int\limits_{-\infty}^\infty \!d\xi \, e^{-i (p+l\kbar/2) \xi/\kbar}
\langle x+\xi/2|\hat U_{\rm free}|\psi_n\rangle\langle\psi_n|\hat U_{\rm free}^\dagger|x-\xi/2\rangle.
\label{eq:wigistep10}
\end{equation}
We can identify the remaining integral in Eq.~(\ref{eq:wigistep10}) when we recall \cite{bib:book} that the free time evolution 
\begin{eqnarray*}
W(x, p; t) &=& \frac{1}{2\pi\kbar}\int\limits_{-\infty}^{\infty} \!d\xi\, e^{-i p \xi/\kbar} \langle x+\frac{1}{2}\xi|\exp\left[-i\frac{\hat p^2}{2\kbar}t\right]|\psi\rangle\langle\psi|\exp\left[i\frac{\hat p^2}{2\kbar}t\right]|x-\frac{1}{2}\xi\rangle
\end{eqnarray*}
of the Wigner function follows from the Wigner function of the initial state $\ket \psi$ by replacing the position $x$ by $x-pt$, that is 
$$
W(x, p; t)=W(x-p t,p; t=0).
$$
Hence, the integral in Eq.~(\ref{eq:wigistep10}) is the Wigner function of the state $\ket{\psi_n}$ propagated for the time $t=1$ and evaluated at the shifted momentum $p+l\kbar/2$. 

Consequently, the recursion formula, Eq.~(\ref{eq:wigistep10}), for the mapping of the Wigner function of the kicked particle takes the form
\begin{equation}
W_{n+1}(x,p) =  \sum_l {\mathcal S}_l(\kappa; x) \, W_n\left(x-\left(p+l\kbar/2\right), p+l\kbar/2\right).
\label{eq:wigmap}
\end{equation}
We recognize that this mapping describes a shearing of the distribution $W_{n+1}$ along the $x$--axis due to the free time evolution. The successive kick causes a displacement in momentum with the position--dependent weight factor $S_l(\kappa; x)$. 

In contrast to the mapping of the state vector, Eq.~(\ref{eq:svmap}) the Wigner function map also involves contributions at half integer multiples of $\kbar$. These additional terms reflect the interference nature of quantum mechanics. They are the analogies of the positive and negative interference structures of a Schr\"odinger cat \cite{bib:pernigo} which are located half way between the classical parts. Moreover, they dissappear when we integrate over position space in order to obtain the momentum distribution. We show this property explicitly in Sec.~\ref{sec:reswig}. 

In the Wigner function treatment of the kicked particle the expansion coefficients ${\mathcal S}_l(\kappa;x)$ play the same role as the coefficients $S_l(\kappa)$ in the state vector description. For a brief comparison between these two quantities and a discussion of their properties we refer to Appendix~\ref{app:sum3}. 

\section{Quantum Resonances viewed from State Space}
\label{sec:resstate}

In this section we use the mapping of the wave function, Eq.~(\ref{eq:svmap}), to discuss a characteristic effect in the time evolution of the kicked particle: the phenomenon of quantum resonances. The size of the scaled Planck's constant $\kbar$ is a decisive factor for the occurrence of quantum effects in the time evolution of the kicked particle. In particular, we obtain dramatically different dynamics for $\kbar$ being a rational and irrational multiple of $4\pi$. In particular, we show that for $\kbar=4\pi$ and integer multiples resonances occur and the average kinetic energy increases quadratically with the number of kicks.

For the value of $\kbar=4\pi$ the mapping defined by the Floquet operator $\hat U$, Eq.~(\ref{eq:ukuf}), simplifies significantly. 
In order to illustrate this interesting dynamics we consider the state 
$$
\ket{\psi_N} = \hat U^N \ket{p=0} = \hat U_{\rm kick} \hat U_{\rm free}\cdot\dots\cdot \hat U_{\rm kick} \hat U_{\rm free}\ket{p=0}
$$
after $N$ kicks where the Floquet operator $\hat U\equiv \hat U_{\rm kick} \hat U_{\rm free}$ acts $N$ times on the initial state $\ket{p=0}$. 

The consequences of the choice $\kbar=4\pi$ stand out most clearly when we consider the free time evolution 
$$
\hat U_{\rm free}\ket{l\kbar} = \exp\left[-i\frac{\hat p^2}{2\kbar}\right]\ket{l\kbar}=\exp\left[-il^2\frac{\kbar}{2}\right]\ket{l\kbar}
$$
of momentum eigenstates $\ket{l\kbar}$ between two kicks. Indeed, for $\kbar=4\pi$ or a integer multiple the phase accumulated by the momentum eigenstate $\ket{l\kbar}=\ket{l\cdot4\pi}$ during the free propagation is $2\pi$ or an integer multiple, that is
\begin{equation}
\hat U_{\rm free}\ket{l\cdot4\pi} = e^{-2\pi i l^2}\ket{l\cdot4\pi} = \ket{l\cdot4\pi}.
\label{eq:ufree}
\end{equation}
Hence, for this particular value of $\kbar$ the momentum eigenstates are invariant under free time evolution.

According to Eq.~(\ref{eq:floquetopmom}) the kick operator $\hat U_{\rm kick}$ couples momentum eigenstates separated by integer multiples of $\kbar$. Since we start from the momentum eigenstate $\ket{p=0}$ we have a superposition of momentum eigenstates $\ket{l\kbar}$ after every kick. For $\kbar=4\pi$ these states are invariant under free time evolution and the state after $N$ kicks reads
$$
\ket{\psi_N} = \hat U_{\rm kick}^N \ket{p=0} = e^{-iN\kappa V(\hat x)}\ket{p=0} =\sum_{l=-\infty}^{\infty}S_l\left(N \kappa\right)\ket{-l\cdot4\pi}
$$
where in the last step we have used the Fourier decomposition, Eq.~(\ref{eq:coestate}), and the shift relation, Eq.~(\ref{eq:shift}). 

With the help of the resulting momentum distribution
\begin{equation}
W_N(p)=\left|\bra p\psi_N\rangle\right|^2=\sum_l S_l^2(N\kappa)\delta(p+l\cdot4\pi),
\label{eq:momstate}
\end{equation}
which involves momenta at $p=l\cdot 4\pi$ with weight factor $S_l^2$, we evaluate the mean energy 
$$
E\equiv\left\langle\textstyle\frac{1}{2}\hat p^2\right\rangle=\int\limits_{-\infty}^\infty \!dp\, \textstyle\frac{1}{2}p^2W_N(p)=8\pi^2\sum_{l=-\infty}^{\infty} l^2 S_l^2(N \kappa)
$$
of the system after $N$ kicks. 

In Appendix~\ref{app:sum1} we calculate this sum over the expansion coefficients $S_l$ and find
$$
E= \frac{1}{2}\langle F^2\rangle K N^2
$$
where $\langle F^2\rangle$ is the square of the force $F=-dV/dx$ averaged over one period. Moreover, we have recalled the abbreviation $\kappa\equiv K/\kbar=K/(4\pi)$.

Hence, for the special choice of $\kbar=4\pi$ the energy increases quadratically with the number of kicks. From Eq.~(\ref{eq:ufree}) we note that also integer multiples of $\kbar=4\pi$ leave the momentum eigenstates invariant. Consequently, the resonance also occurs in these cases. 

We conclude this section by briefly explaining our special choice of the initial state $\ket{p=0}$.
In this case the Floquet operator, Eq.~(\ref{eq:floquetopmom}), only maps this initial state onto discrete momentum eigenstates $p=l\kbar$. During the dynamics we therefore always stay in a momentum ladder starting at zero momentum. Moreover, these eigenstates are invariant under the unitary transformation of the free time evolution provided $\kbar=4\pi$. 

However, there is one more reason for choosing $\ket{p=0}$ as our initial state. Due to the discreteness of the momentum variable the spatial wave function has always a period of $2 \pi$. We would have found the same result if we had imposed periodic boundary conditions corresponding to a kicked rotor. Hence, the mathematics of the kicked particle with zero initial momentum eigenstate and the kicked rotor is identical.

\section{Quantum Resonances viewed from Wigner Phase Space}
\label{sec:reswig}

How does a quantum resonance reflect itself in phase space? In order to answer this question we consider the map, Eq.~(\ref{eq:wigmap}), of the Wigner function.

\subsection{Wigner function after second kick}
\label{sec:wigsec}

The Wigner function of our initial momentum eigenstate $\ket {p=0}$ reads 
$$
W_0(x,p)=\frac{1}{2\pi}\delta(p),
$$ 
where we have introduced the normalization factor $1/(2\pi)$ such that the Wigner function integrated over all momenta and over one spatial period is normalized to unity. 

After one kick, the Wigner function
\begin{equation}
W_{1}(x,p) = \frac{1}{2\pi} \sum_r {\mathcal S}_r(\kappa; x) \, \delta\left(p+r\kbar/2\right).
\label{eq:wig1}
\end{equation}
can be viewed as a stack of delta function walls aligned parallel to the $x$--axis. Each wall is weighted with the function ${\mathcal S}_r$, Eq.~(\ref{eq:wigec}), which imprints a $x$--dependent modulation of period $2\pi$ onto the wall. 

In contrast to the state vector description the Wigner function phase space not only enjoys contributions at $p=l\kbar$ but also at $p=(2l+1)\kbar/2$. However, the corresponding weight function ${\mathcal S}_{2l+1}$ displays a position dependence such that the integral over it vanishes. Only for $p=l\kbar=2l\kbar/2$ do we find a nonvanishing contribution
\begin{equation}
\int\limits_{-\pi}^{\pi} \!dx\,{\mathcal S}_{2l}(\kappa;x)=2\pi S_l^2(\kappa)
\label{eq:relss}
\end{equation}
as shown in Appendix~\ref{app:sum3}. As a consequence, the resulting momentum distribution 
$$
W_1(p)=\int\limits_{-\pi}^{\pi}\!dx\,W_1(x,p)=\sum_l\frac{1}{2\pi}\int\limits_{-\pi}^{\pi}\!dx\, {\mathcal S}_{2l}(\kappa;x)\delta(p+lk)= \sum_lS_l^2(\kappa)\delta(p+l\kbar)
$$
only involves momenta at integer multiples of $\kbar$ in complete agreement with the state vector description.

After the second kick, the phase space distribution
$$
W_2(x,p) = \sum_s {\mathcal S}_s(\kappa; x) \, W_1\left(x-\left(p+s\kbar/2\right), p+s\kbar/2\right).
$$
is expressed in terms of the Wigner function $W_1$, Eq.~(\ref{eq:wig1}), after the first kick which after substitution into this formula yields
$$
W_2(x,p) = \frac{1}{2\pi} \sum_{s} \sum_{r} {\mathcal S}_s(\kappa; x){\mathcal S}_r(\kappa;x-(p+s\kbar/2))\delta\left(p+(r+s)\kbar/2\right).
$$
We first use the $\delta$--function to replace the momentum $p$ in the second expansion coefficient ${\mathcal S}_r$ by $-(r+s)\kbar/2$. We then introduce the summation index $l\equiv r+s$ and arrive at
$$
W_2(x,p) = \frac{1}{2\pi}\sum_{l}{\mathcal W}_l(x)\delta \left(p+l\kbar/2\right).
$$
with the distribution
\begin{equation}
{\mathcal W}_l(x) \equiv\sum_{r} {\mathcal S}_{l-r}(\kappa; x){\mathcal S}_{r}(\kappa;x+r\kbar/2).
\label{eq:wig2}
\end{equation}

\subsection{Resonance}

So far we have not specified the value of $\kbar$. When we now utilize $\kbar=4\pi$ and recognize from the definitions, Eqs.~(\ref{eq:defV}) and (\ref{eq:wigec}), of the generalized potential $\mathcal V$ and the expansion coefficients ${\mathcal S}_l$ the periodicity property ${\mathcal S}_l(\kappa;x+r\cdot 2\pi)={\mathcal S}_l(\kappa;x)$, the coefficient ${\mathcal W}_l$ reduces to 
$$
{\mathcal W}_l^{(+)}\equiv\sum_{r} {\mathcal S}_{l-r}(\kappa; x){\mathcal S}_{r}(\kappa; x) 
$$

In Appendix~\ref{app:sum2} we evaluate this sum analytically and find
$$
{\mathcal W}_l^{(+)} = {\mathcal S}_{l}(2\cdot \kappa; x).
$$
Hence, at a resonance, that is for $\kbar=4\pi$, the phase space distribution after the second kick reads
\begin{equation}
W_2(x,p) =  \frac{1}{2\pi}\sum_r {\mathcal S}_r(2\kappa; x) \, \delta\left(p+r\cdot2\pi\right).
\label{eq:wigsec}
\end{equation}
It is interesting to compare this expression with the phase space distribution
$$
W_1(x,p)=\frac{1}{2\pi}\sum_r {\mathcal S}_r(\kappa;x)\delta(p+r\cdot2\pi)
$$
after the first kick which follows form Eq.~(\ref{eq:wig1}) for $k=4\pi$. We note that the argument $\kappa$ of the expansion coefficient has been replaced by $2\kappa$. 

This result has a simple explanation. During the free time evolution each point of the phase space distribution follows the classical trajectory \cite{bib:book}, that is each point at the momenta $l\kbar/2$ moves with constant velocity and traverses during the time $t=1$ the coordinate distance $x=l\kbar/2\cdot1$. For $\kbar=4\pi$ this distance is one or an integer multiple of $2\pi$. Subsequent to this movement, the next kick occurs. The associated displacement with $x$--dependent weight functions ${\mathcal S}_l(\kappa; x)$ is therefore in phase with the freely propagated phase space distribution and adds up coherently. 

We can continue the iteration of the Wigner function by starting from the distribution, Eq.~(\ref{eq:wigsec}) after the second kick and we find following the above arguments the distribution
$$
{\mathcal W}_l^{(+)}(x) \equiv\sum_{r} {\mathcal S}_{l-r}(\kappa; x){\mathcal S}_{r}(2\kappa;x).
$$
In Appendix~\ref{app:sum2} we have calculated this sum and find the Wigner function
$$
W_3(x,p) =  \frac{1}{2\pi}\sum_r {\mathcal S}_r(3\kappa; x) \, \delta\left(p+r\cdot2\pi\right).
$$
By induction the Wigner function after the $N$-th kick reads
$$
W_N(x,p) =  \frac{1}{2\pi}\sum_r {\mathcal S}_r(N\kappa; x) \, \delta\left(p+r\cdot2\pi\right).
$$

We conclude this section by using this Wigner function to calculate the momentum distribution $W_N(p)$ after $N$ kicks by integrating over position. We recall that this integration over $x$ eliminates the odd momenta, that is
\begin{equation}
W_N(p)=\sum_l\frac{1}{2\pi}\int\limits_{-\pi}^{\pi} \!dx\, {\mathcal S}_{2l}(N\kappa;x)\delta(p+l\cdot4\pi) = \sum_l S_l^2(N\kappa)\delta(p+l\cdot4\pi).
\label{eq:wnres}
\end{equation}
Here we have used the integral relation, Eq.~(\ref{eq:relss}) for the coefficients ${\mathcal S}_{2l}$. 

The result Eq.~(\ref{eq:wnres}) is in complete agreement with the distribution, Eq.~(\ref{eq:momstate}), obtained in the state vector picture.

\subsection{Anti--resonance}

Another interesting case occurs for $\kbar = 2\pi$. This so--called {\it anti-resonance} \cite{bib:izrail,bib:raichl} results from the symmetry of the potential and manifests itself in an oscillating mean energy $E$, that is $E$ oscillates between the initial energy and the energy $\langle F^2\rangle K/2$ after the first kick. Again, we  analyze this phenomenon in phase space. 

According to Eq.~(\ref{eq:wig2}) the distribution ${\mathcal W}_l$ after the second kick reads for $\kbar=2\pi$
$$
{\mathcal W}_l=\sum_{r} {\mathcal S}_{l-r}(\kappa; x){\mathcal S}_{r}(\kappa; x+r\pi).
$$
In Appendix~\ref{app:sum2} we calculate this sum and find ${\mathcal W}_l=\delta_{l,0}$ which yields the Wigner function
$$
W_2(x,p)=\frac{1}{2\pi}\delta(p).
$$

Hence, the Wigner function after two kicks matches exactly the initial Wigner function. Indeed, the free time evolution between the first and second kick has propagated the contributions at the momenta $l\kbar/2$ by $l \pi$ along the $x$ direction. When the second kick occurs, all contributions for the momenta $l\kbar/2$ interfere destructively except for the case $l=0$ where all contributions interfere constructively.

\section{Conclusions and Outlook}
\label{sec:concl}

In the present paper we have studied various aspects of the quantum dynamics of a kicked particle. In Wigner phase space the time evolution is a sequence of shearing the Wigner distribution along the position axis and displacing it along the momentum axis with position dependent weight factors ${\mathcal S}_l(\kappa;x)$. As a first application of this formalism we have revisited the physics of resonances and anti--resonances. However, this approach is also useful when we study the influence of corners in the kicking potential \cite{bib:unpub} on dynamical localization.

Resonances and anti--resonances occur for the special choice $\kbar=4\pi$ and $\kbar=2\pi$, respectively, of the scaled Planck's constant. However, an interesting situation emerges for $\kbar$ being a rational multiple $r/s$ of $4 \pi$. Here the free time evolution shifts each displaced contribution at $p=l\kbar/2$ by an amount of $2\pi l\cdot r/s$. Therefore, consecutive contributions cannot add up fully coherently due to the different spatial shifts of the $x$-dependent weighting functions ${\mathcal S}_l(\kappa; x)$. It is interesting to note that this analytical treatment also includes the limit $r,s\rightarrow\infty$ with $r/s\approx const.$ which corresponds to the limit of localization. Here the contributions of the Wigner function can never interfere coherently and dynamical localization appears, that is the broadening of the momentum distribution is coming to a halt. However, this analysis is beyond the scope of the present paper and will be published elsewhere. 

\section{Acknowledgments}
We thank I.Sh.~Averbukh, B.G.~Englert, S.~Fishman, M.~Freyberger, H.J.~Korsch and Th.~Seligman for many fruitful discussions. This work originated when two of us (FH and WPS) were enjoying the wonderful hospitality of the University of Texas at Austin. We thank our Texan colleagues, in particular D. Steck, for many stimulating discussions during this visit. Moreover, we are most grateful to F.~DeMartini and P.~Mataloni for patiently awaiting the completion of this manuscript. The work of MB and WPS is supported by the Deutsche Forschungsgemeinschaft. MGR gratefully acknowledges the support of the Welch Foundation and the National Science Foundation.

\begin{appendix}

\section{Properties of Wigner Expansion Coefficients}
\label{app:sum3}

In this Appendix we relate the coefficients 
$$
S_l(\kappa)= \frac{1}{2 \pi}\int\limits_{-\pi}^{\pi}\! d \xi \, e^{i l \xi}e^{-i\kappa V(\xi)}
$$ 
emerging from the map of the state vectors to the functions 
\begin{equation}
{\mathcal S}_l(\kappa; x)= \frac{1}{2 \pi} \int\limits_{-\pi}^{\pi} \!dy\, e^{i l y}  e^{-i\kappa V(x+y)}e^{i\kappa V(x-y)}
\label{eq:defsl}
\end{equation}
from the Wigner map. Moreover, we calculate the integral of ${\mathcal S}_l(\kappa;x)$ over one spatial period. This quantity emerging establishes a crucial connection between $S_l(\kappa)$ and ${\mathcal S}_l(\kappa;x)$.

We start the discussion by first noting that both functions are Fourier coefficients. Indeed, the coefficients $S_l$ result from the expansion of $\exp[-i\kappa V(x)]$ whereas ${\mathcal S}_l$ come from $\exp[-i\kappa {\mathcal V}(x,y)]$ where ${\mathcal V}(x,y)\equiv V(x+y)-V(x-y)$. For anti--symmetric potentials $V(-x)=-V(x)$ we find ${\mathcal V}(0,y)=2V(y)$ which establishes the connection
$$
{\mathcal S}_l\left(\kappa; 0\right)=S_l\left(2\kappa\right)
$$
between the two expansion coefficients.

Moreover, it is also important that due to the anti--symmetry $V(-x)=-V(x)$ of the potential and the resulting anti--symmetry ${\mathcal V}(x,-y)=-{\mathcal V}(x,y)$ of the generalized potential both functions $S_l$ and ${\mathcal S}_l$ are purely real. 

The integral
$$
{\mathcal I}_l\equiv \int\limits_{-\pi}^{\pi} \!dx\, {\mathcal S}_l(\kappa;x)
$$
over the expansion coefficients ${\mathcal S}_l(\kappa; x)$ brings out a deep connection with $S_l$. Indeed, when we substitute the expression Eq.~(\ref{eq:defsl}) for ${\mathcal S}_l$ into the definition of ${\mathcal I}_l$ and use the Fourier representations, Eq.~(\ref{eq:expans}), of $\exp[-i\kappa V(x+y)]$ and $\exp[i\kappa V(x-y)]$ we arrive at
$$
{\mathcal I}_l=\sum_r\sum_s S_r S_s\frac{1}{2\pi}\int\limits_{-\pi}^{\pi} \!dx\, e^{-i(r-s)x} \int\limits_{-\pi}^{\pi} \!dy\, e^{i(l-r-s)y}.
$$
The integration over $x$ yields the condition $r=s$ and thus
$$
{\mathcal I}_l=\sum_r S_r^2 \int\limits_{-\pi}^{\pi} \!dy\, e^{i(l-2r)y}.
$$
Hence, we find
$$
{\mathcal I}_{2s+1}=\int\limits_{-\pi}^{\pi} \!dx\, {\mathcal S}_{2s+1}(\kappa;x)=0
$$
and
$$
{\mathcal I}_{2s}=\int\limits_{-\pi}^{\pi} \!dx\, {\mathcal S}_{2s}(\kappa;x)=2\pi S_s^2(\kappa).
$$
For all odd values $l$ the integral of ${\mathcal S}_l$ over one period vanishes, whereas for all even values this integral produces the square of the coefficients $S_l$ of the state space mapping. This feature guarantees that the interference terms in phase space vanish when integrated over position.

\section{Momentum spread}
\label{app:sum1}

In this Appendix we evaluate the sum
$$
I(z)\equiv\sum_{l=-\infty}^{\infty} l^2 S_l^2(z)
$$
containing the expansion coefficients
\begin{equation}
S_l \left(z\right) = \frac{1}{2 \pi}\int\limits_{-\pi}^{\pi} \!d x\, e^{i l x} e^{-izV(x)}
\label{eq:expco}
\end{equation}
of the state vector map. This sum determines the spread of the momentum at the main resonance.

When we substitute the definition, Eq.~(\ref{eq:expco}), of the expansion coefficients $S_l(z)$ into the sum and exchange integration and summation we arrive at
\begin{equation}
I(z)=\frac{1}{4\pi^2}\int\limits_{-\pi}^{\pi} \!dx' \int\limits_{-\pi}^{\pi}\!dx \,
\left[\sum_l l^2 e^{ i l(x+x')}\right] e^{-i z \left[V( x)+V( x')\right]}.
\label{eq:exchsum}
\end{equation}
The representation 
\begin{equation}
\sum_l e^{ i lx}=2\pi\sum_{\nu} \delta(x+2\pi \nu)
\label{eq:comb}
\end{equation}
of a comb of delta functions allows us to express the term in the square brackets in Eq.~(\ref{eq:exchsum}) as the second derivative of the delta function, that is
$$
\sum_l l^2 e^{ i l(x+x')}=-2\pi\sum_{\nu} \frac{\partial^2}{\partial x^2}\delta(x+x'+2\pi \nu).
$$
The derivative can be shifted to the other $x$--dependent part $e^{-i z V(x)}$ of the integrand. When we  integrate over the remaining delta function we find
$$
I(z)=-\frac{1}{2\pi}\int\limits_{-\pi}^{\pi}\!dx \,
 e^{-i z V(-x)}\frac{d^2}{dx^2}e^{-i z V( x)}.
$$
Here we have recognized that the integration only extends over a single interval of $2\pi$ which reduces the summation over $\nu$ to the term $\nu=0$.

The symmetry $V(-x)=-V(x)$ of the potential yields
$$
I(z)=\frac{1}{2\pi} \int\limits_{-\pi}^{\pi}\!dx \left( z^2 \left[\frac{d}{d x}V(x)\right]^2+i z \frac{d^2}{dx^2}V(x)\right).
$$
Due to the anti--symmetry of the potential the second term of the integral does not contribute and we obtain the result
$$
I(z)=\sum_{l=-\infty}^{\infty} l^2 S_l^2(z)=z^2 \cdot \langle F^2\rangle.
$$
Here we have introduced the average
$$
\langle F^2\rangle\equiv\frac{1}{2\pi}\int\limits_{-\pi}^{\pi} \!dx \left[\frac{d}{d x} V(x)\right]^2
$$
of the square of the force $F=-dV/dx$ acting on the particle.

\section{Resonance and Anti--resonance}
\label{app:sum2}

For the phase space analysis of the resonance we need to evaluate the sum
\begin{equation}
{\mathcal W}_l^{(+)}\equiv\sum_{r=-\infty}^{\infty} {\mathcal S}_{l-r}(\kappa; x){\mathcal S}_{r} (\kappa'; x).
\label{eq:defwplus}
\end{equation}
The anti--resonance involves the sum 
\begin{equation}
{\mathcal W}_l^{(-)}\equiv\sum_{r=-\infty}^{\infty} {\mathcal S}_{l-r}(\kappa; x){\mathcal S}_{r}(\kappa; x+r\pi).
\label{eq:defwminus}
\end{equation}

We start our discussion with the sum ${\mathcal W}_l^{(-)}$ and first show that it is closely related to the sum ${\mathcal W}_l^{(+)}$.
For this purpose we introduce the new integration variable $\bar y \equiv -y$ in the expansion coefficient 
\begin{equation}
{\mathcal S}_r(\kappa; x)\equiv \frac{1}{2 \pi} \int\limits_{-\pi}^{\pi} \!dy\,  e^{i r y}  e^{-i\kappa\left[V(x+y)-V(x-y)\right]}
\label{eq:wigexpcoef}
\end{equation}
which yields
$$
{\mathcal S}_{r}(\kappa; x+r\pi)=\frac{1}{2\pi}\int\limits_{-\pi}^{\pi} \!d\bar y\, e^{i (-r) \bar y} e^{-i\kappa\left[V(x+r\pi -\bar y)-V(x+r\pi +\bar y)\right]}.
$$
Due to the periodicity properties $V(x+2m\pi)=V(x)$ and $V(x+(2m+1)\pi)=-V(x)$ of the potential we find the symmetry relation
$$
{\mathcal S}_{r}(\kappa; x+r\pi)={\mathcal S}_{-r}(\kappa; x)
$$
which brings the sum ${\mathcal W}_l^{(-)}$ into the form 
$$
{\mathcal W}_l^{(-)}=\sum_{r=-\infty}^{\infty} {\mathcal S}_{l-r}(\kappa; x){\mathcal S}_{-r}(\kappa; x).
$$

It is therefore convenient to evaluate the sums
$$
{\mathcal C}_l^{(\pm)}\equiv\sum_{r=-\infty}^{\infty} {\mathcal S}_{l-r}(\kappa; x){\mathcal S}_{\pm r}(\kappa'; x).
$$

For this purpose we substitute the definition of ${\mathcal S}_r$, Eq.~(\ref{eq:wigexpcoef}), into the sum and interchange the summation and integration which leads to
$$
{\mathcal C}_l^{(\pm)} =\frac{1}{4\pi^2}\int\limits_{-\pi}^\pi \!dy \int\limits_{-\pi}^\pi \!dy' \left\{\sum_r e^{-i r (y\mp y')}\right\} e^{i l y} e^{-i\left[\kappa{\mathcal V}(x,y) + \kappa'{\mathcal V}(x,y')\right]}.
$$
The sum in the curly brackets represents a comb of delta functions, Eq.~(\ref{eq:comb}), which allows us to perform the integral over $y'$, that is 
\begin{equation}
{\mathcal C}_l^{(\pm)} =\frac{1}{2\pi}\int\limits_{-\pi}^\pi \!dy\, e^{i l y} e^{-i(\kappa\pm\kappa'){\mathcal V}(x,y)}={\mathcal S}_l(\kappa\pm\kappa';x).
\label{eq:cpm}
\end{equation}
Here we have used the symmetry ${\mathcal V}(x,-y)=-{\mathcal V}(x,y)$ following from the definition, Eq.~(\ref{eq:defV}), of ${\mathcal V}$.

Hence, in the sum ${\mathcal W}_l^{(+)}$ defined in Eq.~(\ref{eq:defwplus}) for a resonance the parameters $\kappa$ and $\kappa'$ add. At an anti--resonance the sum ${\mathcal W}_l^{(-)}$, Eq.~(\ref{eq:defwminus}), contains only the single parameter $\kappa$. Since the difference of $\kappa$ and $\kappa'=\kappa$ appears in the explicit expression, Eq.~(\ref{eq:cpm}), for ${\mathcal C}_l^{(-)}$ we find
$$
{\mathcal W}_l^{(-)}= {\mathcal S}_l(0; x)=\frac{1}{2\pi}\int\limits_{-\pi}^\pi \!dy\, e^{i l y}=\delta_{n,0}
$$
where $\delta_{l,0}$ denotes the Kronecker--delta.

\end{appendix}

\end{document}